\newif\ifarxiv
\arxivtrue

\ifarxiv
	\documentclass[12pt]{article}
	\usepackage[margin=1in]{geometry}
	\usepackage[slantedGreek]{mathpazo}
	\usepackage{microtype}
	\usepackage{graphicx}
	\usepackage{amsfonts}
	\usepackage{amssymb}
	\usepackage{amsthm}
	\usepackage{float}
	\usepackage{amsmath}
	\usepackage{hyperref}
	\usepackage{natbib}
	\usepackage{mathrsfs}
	\usepackage{bm}
\else
\fi

\DeclareMathOperator{\Var}{Var}
\DeclareMathOperator{\tr}{tr}
\DeclareMathOperator{\Law}{\mathcal{L}}

\newtheorem{theorem}{Theorem}[section]

\theoremstyle{definition}
\newtheorem{definition}[theorem]{Definition}
\newtheorem{remark}[theorem]{Remark}

\ifarxiv
\title{Bayesian Methods for the Navier-Stokes Equations}
\author{	
	\makebox[.4\linewidth]{Nicholas G. Polson}\\\textit{Booth School of Business}\\\textit{University of Chicago}\\\and 
	\makebox[.4\linewidth]{Vadim Sokolov}\\\textit{Department of Systems Engineering}\\\textit{and Operations Research}\\\textit{George Mason University}
}
\date{First Draft: December 23, 2025\\This Draft: \today}
\else
\fi

\begin{document}
\ifarxiv
\maketitle
\begin{abstract}
We develop a Bayesian methodology for numerical solution of the incompressible Navier--Stokes equations with quantified uncertainty. The central idea is to treat discretized Navier--Stokes dynamics as a state-space model and to view numerical solution as posterior computation: priors encode physical structure and modeling error, and the solver outputs a distribution over states and quantities of interest rather than a single trajectory. In two dimensions, stochastic representations (Feynman--Kac and stochastic characteristics for linear advection--diffusion with prescribed drift) motivate Monte Carlo solvers and provide intuition for uncertainty propagation. In three dimensions, we formulate stochastic Navier--Stokes models and describe particle-based and ensemble-based Bayesian workflows for uncertainty propagation in spectral discretizations. A key computational advantage is that parameter learning can be performed stably via particle learning: marginalization and resample--propagate (one-step smoothing) constructions avoid the weight-collapse that plagues naive sequential importance sampling on static parameters. When partial observations are available, the same machinery supports sequential observational updating as an additional capability. We also discuss non-Gaussian (heavy-tailed) error models based on normal variance-mean mixtures, which yield conditionally Gaussian updates via latent scale augmentation.
\end{abstract}
\else
\fi

\section{Introduction}

The incompressible Navier-Stokes equations,
\begin{equation}\label{eq:ns}
\frac{\partial u}{\partial t} + (u \cdot \nabla)u = -\nabla p + \nu \Delta u + f, \quad \nabla \cdot u = 0,
\end{equation}
govern viscous incompressible fluid motion, where $u$ is velocity, $p$ is pressure, $\nu > 0$ is viscosity, and $f$ is forcing.

This paper develops a Bayesian perspective wherein numerical solution and uncertainty quantification are organized as posterior inference. The Feynman--Kac theorem connects linear parabolic PDEs to expectations over stochastic processes and motivates Monte Carlo solvers in 2D prototype settings; Yudovich \citep{yudovich1963} provides a classical well-posedness benchmark for inviscid 2D vorticity. In 3D, nonlinearity and vortex stretching preclude a direct, closed-form Feynman--Kac reduction, and we focus instead on scalable Bayesian solvers based on particle and ensemble methods. When partial observations are present, these same solvers naturally extend to filtering (data assimilation). In turbulent regimes, heavy-tailed error models (e.g. normal variance-mean mixtures) provide robust observational updates that remain computationally tractable via latent augmentation.

Our goal is to demonstrate how Bayesian modeling and inference can be used to build numerical solvers for Navier--Stokes that quantify uncertainty from discretization, forcing, and modeling error. When partial observations are available, sequential observational updating becomes a natural by-product of the same probabilistic solver rather than its organizing principle. All mathematical material is included only insofar as it supports this numerical Bayesian viewpoint.

Concretely, we:
\begin{itemize}
\item formulate discretized Navier--Stokes as a \textbf{state-space model} (dynamics + observation model) and interpret numerical solution as \textbf{posterior computation};
\item present \textbf{particle} and \textbf{ensemble} workflows for high-dimensional spectral discretizations, highlighting where \emph{bootstrap / full-state} particle filtering degenerates and how practitioners mitigate this (marginalization/Rao--Blackwellization, resample--propagate designs, low-dimensional latent augmentations);
\item highlight \textbf{particle learning} as a practical route to \textbf{online parameter learning without weight collapse}, using sufficient statistics and a resample--propagate (one-step smoothing) construction;
\item use 2D stochastic representations as a \textbf{motivating prototype} for Monte Carlo solvers and uncertainty propagation, emphasizing what is explicit (linear PDE with prescribed drift) versus implicit (fully coupled Navier--Stokes);
\item discuss \textbf{non-Gaussian error models} (heavy tails via normal variance-mean mixtures) that support robust Bayesian updates through latent scale augmentation.
\end{itemize}

Section~\ref{sec:foundations} presents the minimal mathematical background needed for the Bayesian numerical viewpoint. Section~\ref{sec:2d} uses the 2D vorticity setting to motivate stochastic representations and Monte Carlo approximations. Section~\ref{sec:3d} develops particle/ensemble Bayesian solver workflows for 3D discretizations (and shows how observational updates fit when data are available). Section~\ref{sec:robust} discusses non-Gaussian (robust) error models useful in turbulent regimes. Section~\ref{sec:applications} outlines application contexts in geophysics and flow control. Section~\ref{sec:conclusion} concludes.

\section{Mathematical Foundations}\label{sec:foundations}

This section fixes notation and recalls the stochastic and inference objects used throughout (including filtering when observational updates are present). It is intentionally selective: we aim to support Bayesian numerical methodology rather than provide a comprehensive PDE treatment.

\subsection{The Navier-Stokes Equations}

Let $D$ denote either $\mathbb{R}^d$, the periodic torus $\mathbb{T}^d$, or a bounded domain $D \subset \mathbb{R}^d$ with smooth boundary (typically $d = 2$ or $d = 3$). The incompressible Navier-Stokes equations \eqref{eq:ns} express conservation of momentum and mass. The term $\partial u/\partial t$ is local acceleration; $(u \cdot \nabla)u$ is convective acceleration; $-\nabla p$ is the pressure gradient force ensuring incompressibility; $\nu \Delta u$ is viscous diffusion.

Taking the curl of \eqref{eq:ns} eliminates the pressure gradient, yielding the vorticity equation for $\omega = \nabla \times u$:
\begin{equation}\label{eq:vorticity}
\frac{\partial \omega}{\partial t} + (u \cdot \nabla)\omega = (\omega \cdot \nabla)u + \nu \Delta \omega + \nabla \times f.
\end{equation}
In two dimensions, $\omega$ is a scalar (the third component of the curl) and the stretching term $(\omega \cdot \nabla)u$ vanishes identically. In three dimensions, this term represents vortex stretching---the mechanism by which vortex lines intensify as they are stretched by the velocity gradient.

For 2D flows on $\mathbb{R}^2$ (or on the torus with the corresponding periodic kernel), the velocity can be recovered from vorticity via the Biot-Savart law:
\begin{equation}\label{eq:biotsavart}
u(x,t) = \frac{1}{2\pi} \int_{\mathbb{R}^2} \frac{(x-y)^\perp}{|x-y|^2} \omega(y,t) \, dy,
\end{equation}
where $(x_1, x_2)^\perp = (-x_2, x_1)$. Equivalently, introducing the stream function $\psi$ with $u = \nabla^\perp \psi = (\partial_y \psi, -\partial_x \psi)$, the Poisson equation $-\Delta \psi = \omega$ relates stream function and vorticity.

\subsection{Functional Setting for 3D Navier-Stokes}

The mathematical analysis of \eqref{eq:ns} in three dimensions employs the Leray-Hopf theory. For a bounded domain $D \subset \mathbb{R}^3$ with smooth boundary, define
\begin{align}
H &= \text{the closure in $L^2(D)^3$ of }\{u \in C_0^\infty(D)^3 : \nabla \cdot u = 0\}, \\
V &= \{u \in H_0^1(D)^3 : \nabla \cdot u = 0\},
\end{align}
with $V \subset H \subset V'$ forming a Gelfand triple. The Leray-Helmholtz projector $P: L^2(D)^3 \to H$ projects onto divergence-free fields, eliminating the pressure.

\begin{definition}[Leray-Hopf weak solution]
Let $f \in L^2(0,T;V')$ and $u_0 \in H$. A measurable function $u: [0,T] \to H$ is a Leray-Hopf weak solution of \eqref{eq:ns} if:
\begin{enumerate}
\item[(i)] $u \in L^\infty(0,T; H) \cap L^2(0,T; V)$;
\item[(ii)] For all $\phi \in C_0^\infty(D \times [0,T))^3$ with $\nabla \cdot \phi = 0$:
\begin{align}
\int_0^T \int_D \left[ -u \cdot \partial_t \phi - (u \otimes u) : \nabla \phi + \nu \nabla u : \nabla \phi \right] dx \, dt
&= \int_D u_0 \cdot \phi(0) \, dx \\
&\quad + \int_0^T \langle f(t), \phi(t)\rangle_{V',V}\,dt;
\end{align}
\item[(iii)] The energy inequality holds:
\begin{equation}
\frac{1}{2}\|u(t)\|_H^2 + \nu \int_0^t \|\nabla u(s)\|_{L^2}^2 \, ds \leq \frac{1}{2}\|u_0\|_H^2 + \int_0^t \langle f(s), u(s)\rangle_{V',V}\,ds.
\end{equation}
\end{enumerate}
\end{definition}

Leray \citep{leray1934} proved existence of weak solutions in 3D. Uniqueness remains open.

\subsection{Stochastic Calculus}

Let $(\Omega, \mathcal{F}, \{\mathcal{F}_t\}, \mathbb{P})$ be a filtered probability space satisfying the usual conditions. A standard Brownian motion $W = \{W_t\}_{t \geq 0}$ has continuous paths, $W_0 = 0$, independent increments, and $W_t - W_s \sim \mathcal{N}(0, t-s)$.

The It\^o stochastic differential equation
\begin{equation}\label{eq:sde}
dX_t = b(X_t, t) \, dt + \sigma(X_t, t) \, dW_t, \quad X_0 = x_0,
\end{equation}
admits a unique strong solution under Lipschitz and linear growth conditions on $b$ and $\sigma$. The density $p(x,t)$ of $X_t$ (when it exists) satisfies the Fokker-Planck equation
\begin{equation}\label{eq:fp}
\frac{\partial p}{\partial t} = -\nabla \cdot (b p) + \frac{1}{2}\sum_{i,j} \frac{\partial^2}{\partial x_i \partial x_j}(a_{ij} p),
\end{equation}
where $a = \sigma\sigma^\top$ is the diffusion matrix. The backward Kolmogorov equation for $u(x,t) = \mathbb{E}[g(X_T)|X_t = x]$ is
\begin{equation}\label{eq:bk}
\frac{\partial u}{\partial t} + b \cdot \nabla u + \frac{1}{2}\sum_{i,j} a_{ij} \frac{\partial^2 u}{\partial x_i \partial x_j} = 0, \quad u(x,T) = g(x).
\end{equation}

\begin{definition}[Scale mixture of normals]
A random variable $X$ is a scale mixture of normals if $X = \mu + \sigma Z$ where $Z \sim \mathcal{N}(0,1)$ and $\sigma > 0$ is a random variable independent of $Z$. Equivalently, $X|\sigma \sim \mathcal{N}(\mu, \sigma^2)$.
\end{definition}

\begin{remark}
Scale mixtures include the Student-$t$ distribution (inverse-gamma mixing), Laplace distribution (exponential mixing), and Normal Inverse Gaussian distribution (inverse Gaussian mixing).
\end{remark}

\subsection{Bayesian state-space inference (filtering when data are available)}

Consider a diffusion-driven state-space model; when an observation process is present, posterior inference is formulated as filtering:
\begin{align}
dX_t &= f(X_t, t) \, dt + G(X_t, t) \, dW_t, \quad X_0 \sim \pi_0, \label{eq:state}\\
dY_t &= h(X_t, t) \, dt + R^{1/2}(t) \, dV_t, \label{eq:obs}
\end{align}
where $W_t$ and $V_t$ are independent Brownian motions, $X_t \in \mathbb{R}^n$ is the hidden state, and $Y_t \in \mathbb{R}^m$ the observation process.

The filtering problem seeks the conditional distribution
\begin{equation}
\pi_t(dx) = \mathbb{P}(X_t \in dx \mid \mathcal{F}_t^Y), \quad \mathcal{F}_t^Y = \sigma(Y_s : 0 \leq s \leq t).
\end{equation}
Under regularity conditions, $\pi_t$ satisfies the Kushner-Stratonovich equation, a measure-valued SPDE. For nonlinear systems, this equation admits no closed-form solution.

\begin{definition}[Particle filter]
A particle filter approximates $\pi_t$ by a weighted empirical measure:
\begin{equation}
\pi_t^N(dx) = \sum_{i=1}^{N} w_t^i \, \delta_{X_t^i}(dx), \quad \sum_{i=1}^N w_t^i = 1,
\end{equation}
where $\{X_t^i\}_{i=1}^N$ are particle trajectories and $\{w_t^i\}_{i=1}^N$ are importance weights.
\end{definition}

As $N \to \infty$, under suitable conditions, $\pi_t^N \to \pi_t$ weakly. The fundamental Bayesian insight is that the conditional expectation $\mathbb{E}[g(X_t)|\mathcal{F}_t^Y]$ serves as the optimal (minimum variance) estimator of $g(X_t)$.

\section{A 2D prototype: stochastic representations and Monte Carlo solvers}\label{sec:2d}

We use the 2D vorticity formulation as a prototype to connect PDE evolution with stochastic simulation. The key takeaway for numerical methodology is that linear advection--diffusion with prescribed drift admits expectation representations that can be approximated by Monte Carlo, and that the fully coupled Navier--Stokes case can be viewed as an implicit fixed-point version of the same idea.

\subsection{Vorticity Formulation}

In two dimensions, the vorticity equation \eqref{eq:vorticity} reduces to the scalar advection-diffusion equation
\begin{equation}\label{eq:vort2d}
\frac{\partial \omega}{\partial t} + (u \cdot \nabla)\omega = \nu \Delta \omega + (\nabla \times f)_3.
\end{equation}
The velocity $u$ is recovered from $\omega$ via \eqref{eq:biotsavart}. For the viscous case $\nu > 0$, global existence and uniqueness of classical solutions follow from energy estimates.

For the inviscid Euler equations ($\nu = 0$), the situation is more delicate. The vorticity is merely transported: $\frac{D\omega}{Dt} = 0$ along particle paths. Kelvin's circulation theorem states that circulation around material loops is conserved.

\subsection{Yudovich's Uniqueness Theorem}

The fundamental uniqueness result for 2D Euler with rough data is due to Yudovich \citep{yudovich1963}.

\begin{theorem}[Yudovich, 1963]\label{thm:yudovich}
Let $\omega_0 \in L^1(\mathbb{R}^2) \cap L^\infty(\mathbb{R}^2)$. Then the 2D incompressible Euler equations
\begin{equation}\label{eq:euler2d}
\frac{\partial \omega}{\partial t} + (u \cdot \nabla)\omega = 0, \quad u = \nabla^\perp(-\Delta)^{-1}\omega, \quad \omega|_{t=0} = \omega_0,
\end{equation}
admit a unique weak solution $\omega \in L^\infty([0,T]; L^1 \cap L^\infty(\mathbb{R}^2))$ for all $T > 0$.
\end{theorem}

\begin{remark}
The bounded vorticity condition is essentially sharp. For unbounded vorticities, non-uniqueness can occur. Yudovich \citep{yudovich1995} later extended the result to allow $\|\omega_0\|_{L^p}$ to grow at most like $\sqrt{\log p}$ as $p \to \infty$.
\end{remark}

\subsection{The Feynman-Kac Theorem}

The Feynman-Kac formula provides a probabilistic representation of solutions to parabolic PDEs.

\begin{theorem}[Feynman-Kac]\label{thm:fk}
Consider the terminal value problem
\begin{equation}\label{eq:parabolic}
\frac{\partial u}{\partial t} + b(x,t) \cdot \nabla u + \frac{1}{2}\sigma^2(x,t) \Delta u - V(x,t) u = 0, \quad u(x,T) = g(x),
\end{equation}
for $(x,t) \in \mathbb{R}^d \times [0,T]$. Let $\{X_s\}_{s \in [t,T]}$ solve the SDE
\begin{equation}
dX_s = b(X_s, s) \, ds + \sigma(X_s, s) \, dW_s, \quad X_t = x.
\end{equation}
Under suitable regularity conditions on $b$, $\sigma$, $V$, and $g$,
\begin{equation}\label{eq:fk}
u(x,t) = \mathbb{E}\left[ g(X_T) \exp\left(-\int_t^T V(X_s, s) \, ds \right) \Bigg| X_t = x \right].
\end{equation}
\end{theorem}

For the heat equation
\begin{equation}\label{eq:heat}
\frac{\partial \omega}{\partial t} = \nu \Delta \omega, \quad \omega(x,0) = \omega_0(x),
\end{equation}
corresponding to \eqref{eq:vort2d} with prescribed $u = 0$, we have $b = 0$, $\sigma = \sqrt{2\nu}$, $V = 0$. The Feynman-Kac formula (adapted to initial value problems) gives:
\begin{equation}\label{eq:heat-fk}
\omega(x,t) = \mathbb{E}[\omega_0(x + \sqrt{2\nu}W_t)] = \int_{\mathbb{R}^2} G_\nu(x-y,t)\omega_0(y)\,dy,
\end{equation}
where $G_\nu(x,t) = (4\pi\nu t)^{-1}\exp(-|x|^2/(4\nu t))$ is the heat kernel.

The Bayesian interpretation: $\omega(x,t)$ is an expectation over Brownian paths with diffusivity $2\nu$. For a prescribed divergence-free drift field $u$, the linear advection-diffusion equation $\partial_t\omega + u\cdot\nabla\omega = \nu\Delta\omega$ admits a stochastic characteristics representation: if $X_s$ solves the backward SDE
\begin{equation}
dX_s = -u(X_s,s)\,ds + \sqrt{2\nu}\,dW_s,\quad X_t=x,
\end{equation}
then $\omega(x,t)=\mathbb{E}[\omega_0(X_0)]$ (and with forcing, an additional time integral appears). For the full 2D Navier--Stokes system, the drift $u$ is coupled to $\omega$ through the Biot--Savart law, so the representation becomes implicit rather than a closed-form Feynman--Kac formula.

\begin{remark}[Connection to Black--Scholes (computational viewpoint)]
The same ``solution as conditional expectation'' pattern underlies classical option pricing. In the Black--Scholes model \citep{blackscholes1973}, the derivative price can be written as a discounted conditional expectation under a diffusion. For our purposes, the key point is methodological: Feynman--Kac turns a PDE solve into an expectation that can be approximated numerically (Monte Carlo / particle methods), and this is exactly the computational template we exploit for Bayesian numerical Navier--Stokes solvers (and, when data are available, associated observational updates).
\end{remark}

\subsection{Analytical Solutions}

The Feynman-Kac representation yields explicit solutions for certain initial data. These provide useful benchmarks for validating stochastic simulation and Bayesian numerical approximations.

For a point vortex of circulation $\Gamma$ at the origin (the Lamb--Oseen vortex), $\omega_0(x) = \Gamma \delta(x)$:
\begin{equation}
\omega(r,t) = \frac{\Gamma}{4\pi\nu t}\exp\left(-\frac{r^2}{4\nu t}\right), \quad r = |x|.
\end{equation}
The velocity field is
\begin{equation}
u_\theta(r,t) = \frac{\Gamma}{2\pi r}\left(1 - \exp\left(-\frac{r^2}{4\nu t}\right)\right),
\end{equation}
interpolating between solid-body rotation at the core ($u_\theta \sim \Gamma r/(8\pi\nu t)$ as $r \to 0$) and irrotational flow ($u_\theta \sim \Gamma/(2\pi r)$ as $r \to \infty$). The vorticity profile is Gaussian with variance $2\nu t$, matching the variance of Brownian motion.

For two Gaussian vortices of opposite sign separated by distance $d$, the Feynman-Kac representation shows that viscous diffusion eventually causes annihilation, with the separation decreasing due to mutual advection while each vortex core spreads.

\section{Bayesian numerical solvers in three dimensions}\label{sec:3d}

In three dimensions, the primary numerical Bayesian task is \emph{uncertainty-aware solution of a high-dimensional dynamical system}: given a discretized Navier--Stokes model, compute (approximately) a distribution over the evolving velocity field and quantities of interest that reflects discretization, forcing, and modeling error. When partial observations are available, the same probabilistic solver incorporates them through a likelihood and yields filtering as an additional capability. We therefore emphasize stochastic model formulations, discretization, and scalable particle/ensemble algorithms.

\subsection{Motivation: vortex stretching and model uncertainty}

In three dimensions, the vorticity equation \eqref{eq:vorticity} includes the stretching term $(\omega \cdot \nabla)u$, which can amplify vorticity. Consider the decomposition
\begin{equation}
(\omega \cdot \nabla)u = S\omega + \frac{1}{2}\omega \times (\nabla \times u) = S\omega,
\end{equation}
where $S_{ij} = \frac{1}{2}(\partial_j u_i + \partial_i u_j)$ is the strain rate tensor. The vorticity equation becomes
\begin{equation}
\frac{D\omega}{Dt} = S\omega + \nu\Delta\omega.
\end{equation}
When vortex lines align with extensional eigenvectors of $S$, vorticity intensifies. This mechanism drives the energy cascade in turbulent flows and distinguishes 3D from 2D dynamics.

\subsection{Why closed-form stochastic representations are limited in 3D}

The classical Feynman-Kac formula requires a \emph{linear} PDE. The 3D Navier-Stokes equations are nonlinear due to the convective term $(u \cdot \nabla)u$. Three structural features prevent a direct Feynman-Kac representation:

First, nonlinearity implies that the solution $u$ appears in the transport operator, coupling the PDE to the stochastic process that would define the representation.

Second, nonlocality enters through the Biot-Savart law
\begin{equation}\label{eq:biotsavart3d}
u(x,t) = \frac{1}{4\pi} \int_{\mathbb{R}^3} \frac{\omega(y,t) \times (x-y)}{|x-y|^3} \, dy
\end{equation}
means that $u$ at point $x$ depends on $\omega$ everywhere in space.

Third, vortex stretching creates a feedback loop: the stretching term $(\omega \cdot \nabla)u$ creates a feedback loop in which $\omega$ evolves according to an equation involving $\nabla u$, which depends nonlocally on $\omega$ itself.

\subsection{The Constantin-Iyer Representation}

Constantin and Iyer \citep{constantinIyer2008} developed a stochastic Lagrangian formulation that provides a probabilistic representation of Navier--Stokes as an expectation over stochastic flows, though not an explicit Feynman--Kac formula. Consider the stochastic flow map $X(\alpha, t)$ satisfying
\begin{equation}\label{eq:stoch-flow}
dX = u(X,t) \, dt + \sqrt{2\nu} \, dW_t, \quad X(\alpha, 0) = \alpha,
\end{equation}
where $\alpha$ is the Lagrangian label. The velocity admits the representation
\begin{equation}\label{eq:ci-velocity}
u(t,x)=\mathbb{E}\left[\,P\big((\nabla A_t(x))^{\top}u_0(A_t(x))\big)\right],
\end{equation}
where $A_t=X_t^{-1}$ is the (random) inverse flow map and $P$ is the Leray--Helmholtz projector. The dependence of $X_t$ on $u$ makes this an \emph{implicit} fixed-point formulation rather than a closed-form expression.

\begin{remark}
The fundamental difference from Feynman-Kac is summarized as follows:
\begin{center}
\begin{tabular}{ll}
\textbf{Classical Feynman-Kac} & \textbf{Constantin-Iyer} \\
\hline
Linear PDE & Nonlinear PDE \\
Explicit formula & Implicit fixed point \\
Process independent of solution & Process depends on solution \\
Local/semi-local & Nonlocal (Biot-Savart) \\
\end{tabular}
\end{center}
\end{remark}

\subsection{Stochastic Navier-Stokes Equations}

The stochastic Navier-Stokes equations model turbulent fluctuations or forcing uncertainty:
\begin{equation}\label{eq:sns}
du + [\nu Au + B(u,u)] \, dt = Pf \, dt + G \, dW(t),
\end{equation}
where $A = -P\Delta$ is the Stokes operator with domain $D(A) = V \cap H^2(D)^3$, $B(u,v) = P[(u \cdot \nabla)v]$ is the projected nonlinear term, $P$ is the Leray-Helmholtz projector, and $W(t)$ is an $H$-valued Wiener process:
\begin{equation}
W(t) = \sum_{k=1}^{\infty} \sqrt{\lambda_k} \beta_k(t) e_k,
\end{equation}
with $\{\beta_k(t)\}$ independent standard Brownian motions and $\{e_k\}$ an orthonormal basis of $H$.

The mild formulation uses the Stokes semigroup $S(t) = e^{-\nu tA}$:
\begin{equation}\label{eq:mild}
u(t) = S(t)u_0 - \int_0^t S(t-s)B(u(s),u(s)) \, ds + \int_0^t S(t-s)Pf \, ds + \int_0^t S(t-s)G \, dW(s).
\end{equation}

\begin{theorem}[\citet{flandoli1994}]
Let $D \subset \mathbb{R}^3$ be bounded with smooth boundary, $u_0 \in H$, and assume trace-class noise: $\tr(GG^*) = \sum_k \lambda_k < \infty$. Then:
\begin{enumerate}
\item[(i)] There exists a martingale solution of \eqref{eq:sns}.
\item[(ii)] In 2D, the solution is unique.
\item[(iii)] In 3D, uniqueness remains open.
\end{enumerate}
\end{theorem}

\subsection{Interacting Particles and McKean-Vlasov Equations}

The Navier-Stokes equations (especially in vorticity form) motivate interacting particle and mean-field approximations \citep{mckean1966, kurtzXiong1999}. A canonical McKean--Vlasov particle system has $N$ particles $X_i(t) \in \mathbb{R}^d$:
\begin{equation}\label{eq:particles}
dX_i = (K * \mu_t^N)(X_i)\,dt + \sqrt{2\nu} \, dW_i,
\end{equation}
where $\mu_t^N = \frac{1}{N}\sum_{j=1}^N \delta_{X_j(t)}$, $K$ is an interaction kernel, and $\{W_i\}$ are independent Brownian motions. The empirical measure
\begin{equation}
\mu^N_t = \frac{1}{N}\sum_{i=1}^{N} \delta_{X_i(t)}
\end{equation}
satisfies a stochastic PDE. As $N \to \infty$, propagation of chaos implies that particles become asymptotically independent, each with law $\mu_t$ satisfying the McKean-Vlasov equation:
\begin{equation}\label{eq:mckean}
dX_t = b(X_t, \mu_t) \, dt + \sigma(X_t, \mu_t) \, dW_t, \quad \mu_t = \Law(X_t).
\end{equation}

For vortex methods in 2D, the kernel is the Biot-Savart kernel $K(x) = x^\perp/(2\pi|x|^2)$, and the limit recovers the vorticity equation \eqref{eq:vort2d} \citep{cottetKoumoutsakos2000}. In 3D, the situation is more complex due to vortex stretching, but analogous constructions exist using vortex filament or vortex blob methods.

\subsection{Mean-Field Limits and Fokker-Planck Connection}

The mean-field limit $N\to\infty$ connects interacting particle systems to nonlinear Fokker--Planck equations for $\mu_t$, and in vorticity-based settings provides a natural route to particle approximations of Navier--Stokes and related filtering problems.

\subsection{Spectral Discretization}

For computational purposes, we discretize in a Fourier basis. On the periodic domain $D = [0,2\pi]^3$:
\begin{equation}
u(x,t) = \sum_{k \in \mathbb{Z}^3} \hat{u}_k(t) e^{ik \cdot x}, \quad k \cdot \hat{u}_k = 0 \text{ (incompressibility)}.
\end{equation}
The spectral Navier-Stokes equations are
\begin{equation}
\frac{d\hat{u}_k}{dt} = -\nu|k|^2 \hat{u}_k - \sum_{j+m=k} P_k\!\left[i(m \cdot \hat{u}_j)\hat{u}_m\right] + \hat{f}_k + \hat{G}_k \dot{W}_k,
\end{equation}
where $P_k = I - kk^T/|k|^2$ is the Leray projector in Fourier space.

Truncating to $|k| \leq K_{\max}$ yields a finite-dimensional system with $M = O(K_{\max}^3)$ degrees of freedom, suitable for particle and ensemble Bayesian solvers. When such a solver is coupled to observations, it becomes a filtering method.

\subsection{Particle-based Bayesian solvers (with optional observations)}

The particle-filtering formalism can be used as a \emph{probabilistic numerical solver}: it propagates an ensemble of weighted trajectories for the discretized dynamics and thereby represents uncertainty in the evolving state. When partial observations $y_k$ are available, the same machinery incorporates them through a likelihood and produces an updated posterior distribution of the velocity field. For the discretized system with state $u \in \mathbb{R}^{2M}$ and (optional) observations $y_k$ at times $t_k$, we use the state equation

$u_{k+1} = u_k + \Delta t \, F(u_k) + \sqrt{\Delta t} \, G\xi_k$ with $\xi_k \sim \mathcal{N}(0,I)$, and the observation equation $y_k = Hu_k + \eta_k$ with $\eta_k \sim \mathcal{N}(0,R)$.

The bootstrap particle filter maintains particles $\{u_k^i\}_{i=1}^N$ and weights $\{w_k^i\}$:
\begin{enumerate}
\item \textbf{Prediction:} $u_{k+1}^i = u_k^i + \Delta t \, F(u_k^i) + \sqrt{\Delta t} \, G\xi_k^i$.
\item \textbf{Update:} $\tilde{w}_{k+1}^i = w_k^i \cdot p(y_{k+1}|u_{k+1}^i)$, then normalize.
\item \textbf{Resample:} If $N_{\text{eff}} = 1/\sum_i(w^i)^2 < N_{\text{thresh}}$, resample.
\end{enumerate}

The posterior mean is $\bar{u}_k = \sum_i w_k^i u_k^i$. Particle learning methods \citep{carvalhoEtAl2010} extend this framework by propagating \emph{sufficient statistics} for unknown parameters and by using \emph{marginalization/resample--propagate} constructions (auxiliary particle filters) that select ancestors using one-step-ahead information. This largely avoids the weight-degeneracy pathology associated with naive parameter learning via sequential importance sampling, while keeping the update local in time.

For very high-dimensional states ($M \gtrsim 10^4$), \emph{full-state bootstrap particle filtering} suffers weight degeneracy unless additional structure is exploited. The ensemble Kalman filter (EnKF) provides an alternative using Gaussian approximations with localization to suppress spurious long-range correlations \citep{evensen2009, reich2015}.

\section{Non-Gaussian Error Models for Robust Bayesian Solvers}\label{sec:robust}

\subsection{Motivation: heavy tails and robustness}

In turbulent regimes, innovations and observation errors are often heavy-tailed, and Gaussian likelihoods can yield brittle filters (weight collapse in particle filters; sensitivity to outliers in Gaussian updates). A pragmatic Bayesian remedy is to use heavy-tailed likelihoods that retain computational tractability via conditional-Gaussian augmentation.

\subsection{Normal variance-mean mixtures}

Normal variance-mean mixtures provide a broad class of heavy-tailed models while preserving conditional Gaussian structure. In the scalar case, if
\begin{equation}\label{eq:nvm-scalar}
X \mid \tau \sim \mathcal{N}(\mu + \beta\tau,\ \tau), \qquad \tau>0,
\end{equation}
then marginally $X$ is heavy-tailed whenever $\tau$ has sufficient mass near $0$ and/or heavy tails. In the multivariate case one uses
\begin{equation}\label{eq:nvm-vector}
X \mid \tau \sim \mathcal{N}(\mu + \beta\tau,\ \tau \Sigma),
\end{equation}
where $\Sigma$ is a base covariance and $\tau$ is a positive scalar (or, for component-wise robustness, a vector of independent scales).

The key computational benefit is that inference can be performed by augmenting $\tau$. Conditional on $\tau$, one recovers a Gaussian likelihood/dynamics, so Kalman updates and Gaussian-weight computations remain available. This is especially useful in high-dimensional Navier--Stokes settings where one wants robust observational updates without abandoning scalable Gaussian computations.

\subsection{Inverse-gamma vs inverse-Gaussian mixing (terminology)}

The literature contains two closely related but distinct ``normal--inverse-\(\cdot\)'' constructions that are easy to confuse:
\begin{itemize}
\item \textbf{Inverse-gamma mixing (Student-$t$).} If $\tau \sim \text{Inv-Gamma}(\cdot)$ and $X \mid \tau$ is Gaussian with variance proportional to $\tau$, then $X$ is Student-$t$. This yields very heavy (polynomial) tails and is a common robust alternative to Gaussian likelihoods.
\item \textbf{Inverse-Gaussian mixing (Normal Inverse Gaussian).} If $\tau$ is inverse Gaussian and $X \mid \tau$ is Gaussian with mean/variance affine in $\tau$, then $X$ is Normal Inverse Gaussian (NIG). This yields semi-heavy (exponential) tails and, crucially for sequential computation, leads to convenient conditional distributions for $\tau$.
\end{itemize}

In what follows, ``NIG'' always means \emph{Normal Inverse Gaussian} (inverse-\emph{Gaussian} mixing).

\subsection{Normal Inverse Gaussian distribution}

The Normal Inverse Gaussian (NIG) distribution \citep{barndorffnielsen1997} is a canonical normal variance-mean mixture \citep{barndorffnielsenKentSorensen1982}. It is a convenient choice for robust likelihoods because it admits a latent inverse-Gaussian scale representation.

We use the common parameterization $\text{NIG}(\alpha,\beta,\mu,\delta)$ with constraints $\delta>0$ and $\alpha>|\beta|$. Let $\gamma=\sqrt{\alpha^2-\beta^2}$.

\begin{definition}
$X \sim \text{NIG}(\alpha,\beta,\mu,\delta)$ if
\begin{equation}
X = \mu + \beta \tau + \sqrt{\tau} Z,
\end{equation}
where $Z \sim \mathcal{N}(0,1)$, $Z \perp \tau$, $\gamma=\sqrt{\alpha^2-\beta^2}$, and $\tau$ is inverse Gaussian with mean $\delta/\gamma$ (equivalently, $\tau\sim\text{IG}(\delta/\gamma,\delta^2)$ under the common mean/shape parametrization).
\end{definition}

The NIG density is
\begin{equation}
f(x; \alpha,\beta,\mu,\delta) = \frac{\alpha\delta}{\pi} e^{\delta\gamma + \beta(x-\mu)} \frac{K_1(\alpha q(x))}{q(x)},
\end{equation}
where $\gamma = \sqrt{\alpha^2-\beta^2}$, $q(x) = \sqrt{\delta^2 + (x-\mu)^2}$, and $K_1$ is a modified Bessel function.

The characteristic function is
\begin{equation}\label{eq:nig-cf}
\varphi_X(t)=\mathbb{E}[e^{itX}]
=\exp\!\left(i\mu t + \delta\left(\gamma-\sqrt{\alpha^2-(\beta+it)^2}\right)\right).
\end{equation}
Differentiating \eqref{eq:nig-cf} yields
\begin{align}
\mathbb{E}[X] &= \mu + \delta \frac{\beta}{\gamma}, \\
\Var(X) &= \delta \frac{\alpha^2}{\gamma^3}.
\end{align}
The pair $(\alpha,\beta)$ controls tail thickness and skewness; $\delta$ is a scale parameter; $\mu$ is a location parameter.

Properties relevant for robust Bayesian updating:
\begin{enumerate}
\item[(i)] Scale mixture of normals with inverse Gaussian mixing.
\item[(ii)] Semi-heavy tails: $f(x) \sim |x|^{-3/2}e^{-\alpha|x|}$ as $|x| \to \infty$.
\item[(iii)] Closed under convolution: if $X_i \sim \text{NIG}(\alpha,\beta,\mu_i,\delta_i)$ independently, then $\sum X_i \sim \text{NIG}(\alpha,\beta,\sum\mu_i,\sum\delta_i)$.
\item[(iv)] Heavy tails and skewness can be represented while retaining conditional Gaussian structure.
\end{enumerate}

\subsection{Latent-scale augmentation and conditional posteriors}

For Bayesian computation, the key object is the conditional distribution of $\tau$ given $X=x$ under the mixture representation. From
\[
x-\mu = \beta\tau + \sqrt{\tau}z, \qquad z\sim\mathcal{N}(0,1),
\]
we have
\begin{equation}
p(x\mid \tau)\ \propto\ \tau^{-1/2}\exp\!\left(-\frac{(x-\mu-\beta\tau)^2}{2\tau}\right).
\end{equation}
The inverse-Gaussian prior on $\tau$ implied by the NIG parameterization can be written in exponential form as
\begin{equation}
p(\tau)\ \propto\ \tau^{-3/2}\exp\!\left(-\frac{1}{2}\left(\frac{\delta^2}{\tau}+\gamma^2\tau\right)\right),
\end{equation}
up to a normalizing constant. Multiplying yields the conditional posterior
\begin{equation}\label{eq:tau-given-x}
p(\tau\mid x)\ \propto\ \tau^{-3/2}\exp\!\left(-\frac{1}{2}\left(\frac{\delta^2+(x-\mu)^2}{\tau}+\alpha^2\tau\right)\right),
\end{equation}
which is a \emph{generalized inverse Gaussian} (GIG) distribution with index $\lambda=-\tfrac{1}{2}$, ``inverse-scale'' parameter $\chi=\delta^2+(x-\mu)^2$, and ``scale'' parameter $\psi=\alpha^2$:
\[
\tau\mid x \sim \mathrm{GIG}\!\left(\lambda=-\tfrac{1}{2},\ \chi=\delta^2+(x-\mu)^2,\ \psi=\alpha^2\right).
\]

Equation \eqref{eq:tau-given-x} is the workhorse for Bayesian algorithms: one can sample $\tau$ (e.g. via standard GIG samplers) or work with conditional expectations such as $\mathbb{E}[\tau\mid x]$ and $\mathbb{E}[\tau^{-1}\mid x]$ in EM/variational approximations. The resulting computations plug directly into filtering updates because conditional on $\tau$ the model is Gaussian.

\subsection{NIG likelihoods for robust observational coupling}

To robustify the observation model $y_k=H u_k+\eta_k$, replace Gaussian observation noise by a component-wise NIG model. One convenient specification is
\begin{align}
y_k \mid u_k,\tau_k &\sim \mathcal{N}(H u_k + \beta \tau_k,\ \tau_k R), \label{eq:nig-obs-aug}\\
\tau_k &\sim \text{IG}(\delta/\gamma,\delta^2), \nonumber
\end{align}
where $\tau_k$ may be taken scalar (global scale at time $k$) or vector-valued (independent scales per sensor/per Fourier mode) to target localized outliers. Marginally, each component is NIG.

The latent $\tau_k$ acts as a \emph{random inflation} of observation covariance: outliers are explained by large $\tau_k$ rather than forcing the filter to overreact, yielding robustness.

Given a current predicted mean $\hat{u}_{k|k-1}$, the (scalar) innovation is $e_k = y_k - H\hat{u}_{k|k-1}$. In the simplest scalar-noise case, the conditional law $\tau_k\mid e_k$ is again GIG of the form \eqref{eq:tau-given-x} with $x\leftarrow e_k$ (and with appropriate scaling if $R\neq 1$). Conditional on $\tau_k$, the update is Gaussian with covariance $\tau_k R$ and mean shift $\beta\tau_k$.

\subsection{Algorithms: detailed robust filters}

We now spell out the algorithmic patterns that (to our knowledge) are rarely written explicitly in the Navier--Stokes numerical literature, despite being straightforward once the augmentation is made explicit.

For particle filtering with NIG observations, assume the dynamics are approximated by a (possibly nonlinear) forecast step $u_{k+1}=F(u_k)+\varepsilon_k$. For simplicity, take the observation model \eqref{eq:nig-obs-aug} with scalar $\tau_k$ and define the innovation for particle $i$ as
\[
e_k^{(i)} = y_k - H u_{k|k-1}^{(i)}.
\]
The particle filter is run on an \emph{extended state} $(u_k,\tau_k)$.

\emph{(A) Prior proposal for $\tau_k$ (Bessel-free).}
Propagate $\tau_k$ from its prior and weight using the conditional Gaussian likelihood:
\begin{enumerate}
\item \textbf{Forecast:} propagate particles $u_{k|k-1}^{(i)} \sim p(u_k\mid u_{k-1}^{(i)})$ using the numerical time-stepper (plus any stochastic model error).
\item \textbf{Scale propagation:} sample $\tau_k^{(i)} \sim p(\tau_k)$ from the inverse-Gaussian prior in \eqref{eq:nig-obs-aug}.
\item \textbf{Weight update:} evaluate the Gaussian likelihood
\[
p(y_k\mid u_{k|k-1}^{(i)},\tau_k^{(i)})=\mathcal{N}\!\left(y_k;\ H u_{k|k-1}^{(i)}+\beta\tau_k^{(i)},\ \tau_k^{(i)}R\right),
\]
and update weights
\[
w_k^{(i)} \propto w_{k-1}^{(i)}\,p(y_k\mid u_{k|k-1}^{(i)},\tau_k^{(i)}).
\]
\item \textbf{Resample (if needed):} resample particles when $N_{\mathrm{eff}}$ falls below a threshold.
\end{enumerate}
This variant avoids evaluation of modified Bessel functions and is often the most practical starting point.

\emph{(B) Fully adapted proposal for $\tau_k$ (lower variance, needs marginal).}
Alternatively, propose $\tau_k^{(i)} \sim p(\tau_k\mid e_k^{(i)})$ using the GIG form \eqref{eq:tau-given-x} (with scaling by $R$ as needed). In this case the incremental importance weight becomes the \emph{marginal} likelihood
\[
w_k^{(i)} \propto w_{k-1}^{(i)}\,p(y_k\mid u_{k|k-1}^{(i)}),
\]
where $p(y_k\mid u)$ is the NIG density (or its multivariate analogue under independence assumptions). This proposal typically reduces weight variance but requires evaluating the NIG marginal (involving $K_1$) or approximating it.

In ensemble Kalman filtering, Gaussianity is assumed only at the level of mean/covariance updates; a practical robust variant is the scale-mixture EnKF (SM-EnKF):
\begin{enumerate}
\item draw an ensemble forecast $\{u_{k|k-1}^{(j)}\}_{j=1}^N$ by propagating the numerical solver;
\item for each observation (or for a global scale) sample a scale $\tau_k$ from its conditional or from a prior predictive robust rule based on innovations;
\item perform the EnKF analysis step using the inflated observation covariance $\tau_k R$ (and mean shift $\beta\tau_k$ if included).
\end{enumerate}
This retains EnKF scalability while providing heavy-tailed robustness through random covariance inflation.

\subsection{Particle learning with NIG augmentation (resample--propagate / one-step smoothing)}

Particle learning \citep{carvalhoEtAl2010} augments particle methods with \emph{parameter learning} by propagating low-dimensional sufficient statistics and by using a resample--propagate construction that is close to an auxiliary particle filter. Conceptually, particles are first \emph{selected} using one-step-ahead information (a one-step smoothing/ancestor selection step), and only then \emph{propagated} forward. This is the key reason particle learning does not exhibit the same ``degenerate weights'' behavior as naive sequential importance sampling on static parameters.

In practice, particle learning is often implemented by marginalizing (integrating) analytically tractable blocks and propagating only the remaining low-dimensional uncertainty with particles. In our setting, NIG models are particularly compatible with this strategy because conditioning on latent scales restores Gaussian structure.

Consider a state-space model with NIG innovations:
\begin{align}
u_t | u_{t-1}, \tau_t &\sim \mathcal{N}(f(u_{t-1}), \tau_t Q), \label{eq:nig-state}\\
y_t | u_t, \tau_t &\sim \mathcal{N}(H u_t, \tau_t R), \\
\tau_t &\sim \text{IG}(\delta/\gamma, \delta^2),
\end{align}
where $\tau_t$ is the latent inverse Gaussian scale variable. Marginalizing over $\tau_t$, the innovations follow NIG distributions, capturing the heavy tails observed in turbulent velocity increments.

In high-dimensional Navier--Stokes settings, one typically treats the velocity state $u_t$ as high-dimensional and the scales $\tau_t$ (and parameters) as low-dimensional. A common and effective design is therefore:
\begin{itemize}
\item sample $\tau_t$ (and possibly parameters) with SMC;
\item condition on $\tau_t$ to perform Gaussian computations for $u_t$ (Kalman-type updates, or Gaussian likelihood evaluations for weights).
\end{itemize}
This is exactly the marginalization/Rao--Blackwellization principle: keep particles for the low-dimensional non-Gaussian part and integrate the rest analytically when possible. When combined with resample--propagate (auxiliary) designs, this yields stable sequential learning of parameters and latent scales without confusing it with the degeneracy of full-state bootstrap particle filters in very high dimension.

To learn $(\mu,\beta)$ in the NIG observation model, consider a scalar residual model $r_t = y_t - H u_t$ with
\[
r_t \mid \tau_t,\mu,\beta \sim \mathcal{N}(\mu+\beta\tau_t,\ \tau_t),
\]
define $\theta=(\mu,\beta)^\top$ and the (random) design vector $x_t=(1,\tau_t)^\top$. Conditional on $\tau_{1:t}$, this is weighted linear regression with weight $1/\tau_t$. With a Gaussian prior $\theta\sim\mathcal{N}(m_0,V_0)$, the posterior is Gaussian with updated sufficient statistics
\begin{align}
A_t &= \sum_{s=1}^t \frac{1}{\tau_s} x_s x_s^\top
=
\begin{pmatrix}
\sum_{s=1}^t \tau_s^{-1} & \sum_{s=1}^t 1 \\
\sum_{s=1}^t 1 & \sum_{s=1}^t \tau_s
\end{pmatrix},\\
b_t &= \sum_{s=1}^t \frac{1}{\tau_s} x_s r_s
=
\begin{pmatrix}
\sum_{s=1}^t r_s/\tau_s \\
\sum_{s=1}^t r_s
\end{pmatrix}.
\end{align}
Then
\[
V_t^{-1}=V_0^{-1}+A_t,\qquad m_t=V_t\left(V_0^{-1}m_0+b_t\right),
\]
and particle learning proceeds by carrying $(A_t,b_t)$ (or equivalently the scalar sums $\sum \tau_s^{-1},\sum\tau_s,\sum r_s,\sum r_s/\tau_s$) inside each particle. The same idea extends component-wise to vector observations or per-mode residuals.

To learn the NIG tail parameters $(\alpha,\delta)$ (tail/scale) from augmented scales, note that unlike $(\mu,\beta)$ these enter through the prior on $\tau_t$ and are not conjugate in general. However, conditional on sampled scales $\tau_{1:t}$, the log-likelihood contributions depend on low-dimensional summaries such as $\sum \tau_s$ and $\sum \tau_s^{-1}$ (up to normalizing constants involving modified Bessel functions). In practice one can:
\begin{itemize}
\item fix $(\alpha,\delta)$ based on a calibration window and then run online filtering, or
\item embed a numerical M-step (or occasional MCMC refresh) that updates $(\alpha,\delta)$ using the accumulated scale samples within particles.
\end{itemize}

The variance reduction from Rao--Blackwellization is substantial. By analytically integrating over the Gaussian state $u_t$ via Kalman filtering, particles need only track the lower-dimensional scale variable $\tau_t$. This reduces the effective variance by a factor proportional to the state dimension, making the approach tractable for high-dimensional turbulent flow fields.

In this particle-learning view, NIG provides a robust, heavy-tailed alternative to Gaussian errors while keeping the \emph{sequential computations} close to standard Gaussian filtering: the only new moving parts are (i) sampling or approximating the latent scales and (ii) optionally propagating compact sufficient statistics for the parameters.

\subsection{Numerical demonstration: particle learning avoids weight collapse (parameter learning)}
To make the particle-learning advantage concrete, we include a minimal toy experiment comparing particle learning to naive sequential importance sampling (SIS) for learning a static parameter from a stream of observations. The SIS approach multiplies likelihoods over time without rejuvenation and therefore exhibits severe weight collapse, whereas particle learning uses sufficient statistics and a resample--propagate (one-step smoothing) construction that keeps the particle population diverse while tracking the posterior over parameters.

\begin{figure}[H]
\centering
\includegraphics[width=0.85\linewidth]{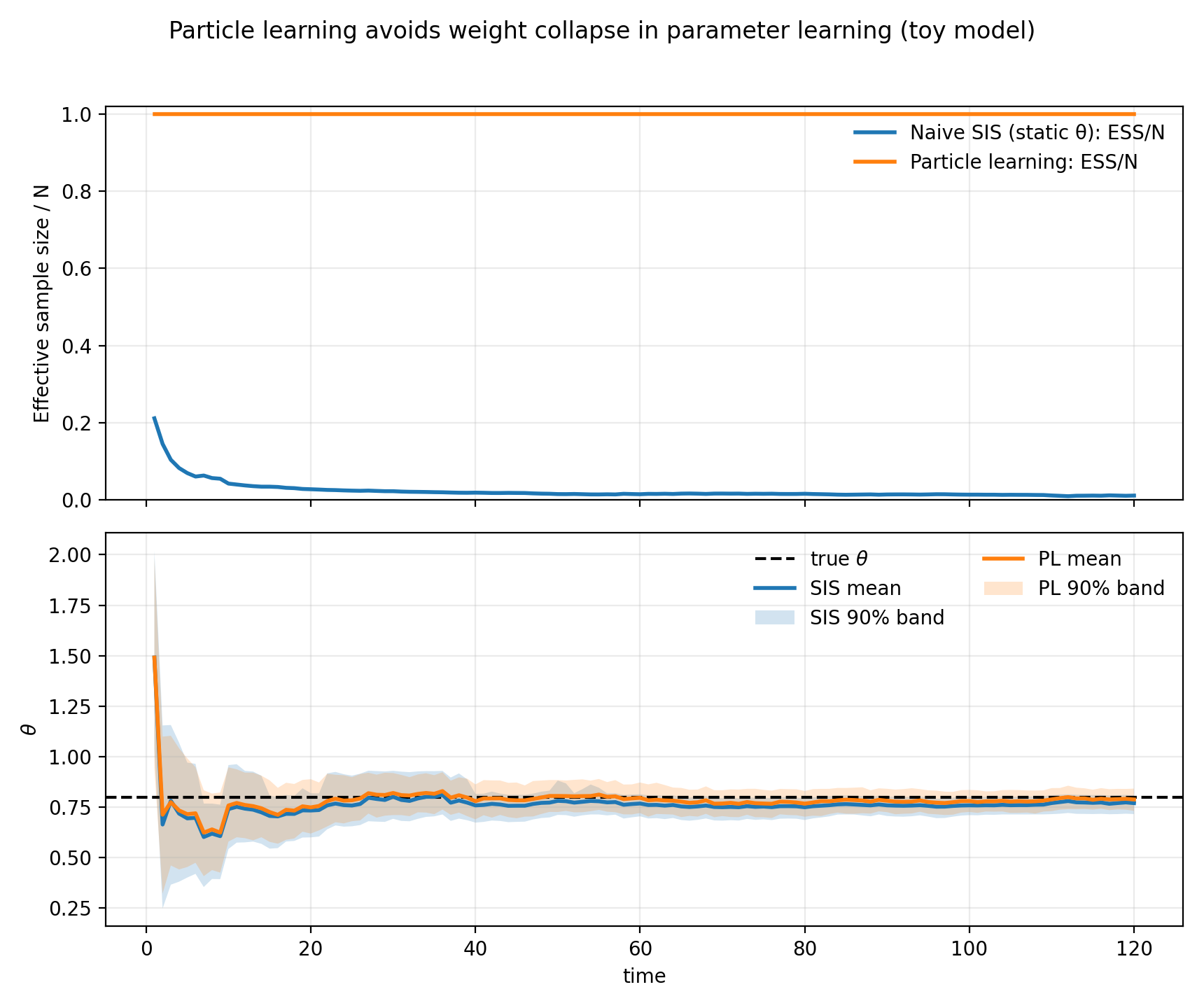}
\caption{Toy parameter-learning example illustrating a key computational advantage of particle learning. Naive sequential importance sampling (SIS) on a static parameter suffers rapid weight collapse (ESS decays), while particle learning maintains ESS by combining sufficient statistics (marginalization) with a resample--propagate (one-step smoothing) update.}
\label{fig:pl-vs-sis}
\end{figure}

Figure~\ref{fig:pl-vs-sis} shows the effective sample size (ESS) over time and the resulting posterior bands for the parameter. The script \texttt{code/demo\_particle\_learning\_vs\_sis.py} reproduces the figure.

\subsection{Numerical demonstration: a particle solver for 2D Navier--Stokes (no observations)}
To illustrate the solver viewpoint directly (without assimilation), we simulate the 2D Navier--Stokes equations in vorticity form on the periodic torus using a pseudo-spectral semi-implicit time stepper. We place a prior on uncertain physical/model parameters---here viscosity $\nu$ and forcing amplitude---and propagate this prior through the numerical solver using a particle/ensemble approximation. The output is a posterior predictive distribution over solution fields and quantities of interest (e.g. kinetic energy and enstrophy), obtained purely from uncertainty propagation through the PDE solver.

\begin{figure}[H]
\centering
\includegraphics[width=0.95\linewidth]{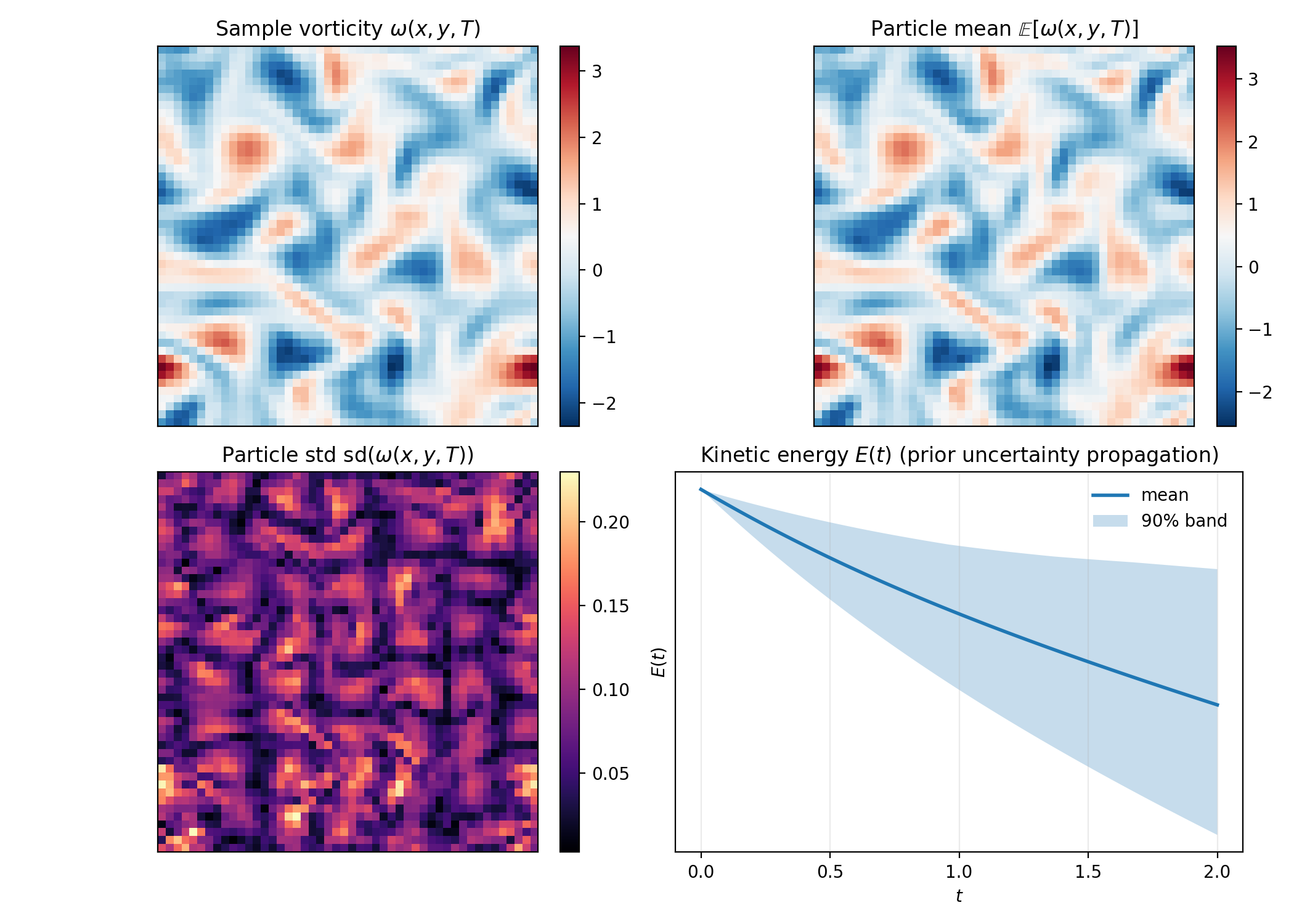}
\caption{Particle/ensemble Bayesian numerical solution of 2D Navier--Stokes (vorticity form) on a torus with uncertain viscosity and forcing amplitude. No observations are used: uncertainty is propagated through the solver to produce distributions over the vorticity field and kinetic energy.}
\label{fig:ns2d-solver-uq}
\end{figure}

Figure~\ref{fig:ns2d-solver-uq} shows a representative run: sample/mean/standard deviation of the terminal vorticity field, together with a 90\% uncertainty band for kinetic energy. The script \texttt{code/demo\_ns2d\_particle\_solver.py} reproduces the figure and prints summary statistics.

\subsection{Numerical demonstration: robust observational updates under outliers}

To illustrate why latent-scale models matter computationally, we simulate a 1D linear state-space model with occasional observation outliers and compare a mis-specified Gaussian Kalman filter to an NIG-style scale-augmented Rao--Blackwellized particle filter (prior proposal for scales, Gaussian conditional likelihood). In this toy example, the RBPF reduces the RMSE and improves uncertainty calibration (90\% interval coverage) under outliers.

\begin{figure}[H]
\centering
\includegraphics[width=0.95\linewidth]{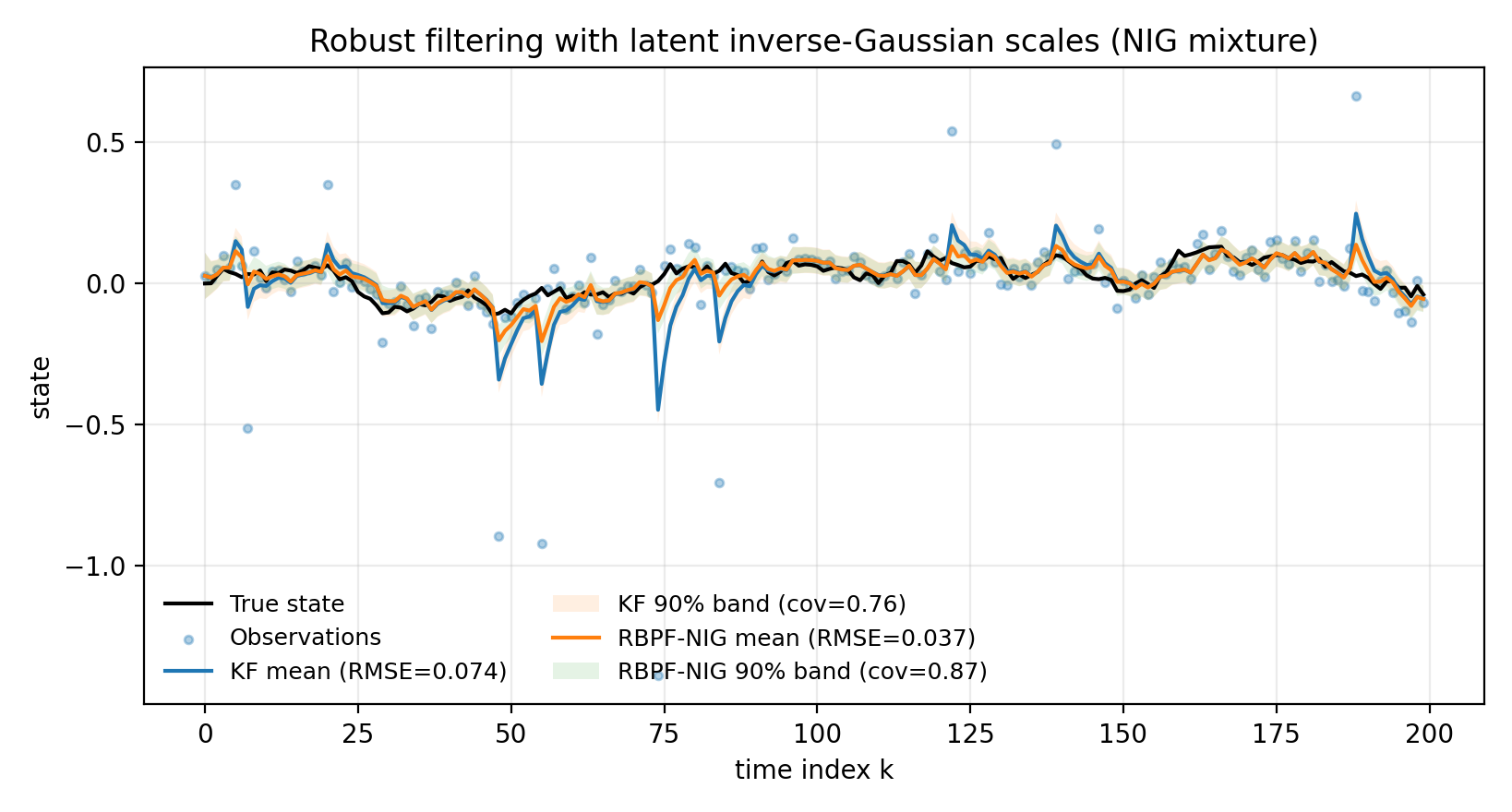}
\caption{Gaussian Kalman filter vs scale-augmented (NIG-motivated) Rao--Blackwellized particle filter in a 1D linear system with outliers. Heavy-tailed robustness is achieved by sampling latent inverse-Gaussian scales that inflate observation variance during outlier events.}
\label{fig:nig-rbpf}
\end{figure}

Figure~\ref{fig:nig-rbpf} shows a representative run. The script \texttt{code/demo\_nig\_rbpf\_1d.py} reproduces the plot and prints summary metrics (for the shown run: RMSE $0.074$ for KF vs $0.037$ for RBPF; 90\% coverage $0.76$ for KF vs $0.87$ for RBPF).

\section{Applications}\label{sec:applications}

\subsection{How the Bayesian solver is instantiated in practice}

This section translates the methodology into a concrete modeling and computation recipe. In each application, we begin with a spatial discretization (spectral / finite volume / finite element) that yields a finite-dimensional state $u_k \in \mathbb{R}^d$, specify a stochastic dynamical model $u_{k+1}=F_k(u_k)+\xi_k$ that encodes numerical time stepping, subgrid effects, and model error, and couple it to partial observations through an observation model $y_k = H_k u_k + \eta_k$ (allowing non-Gaussian $\eta_k$ when appropriate). Given these ingredients, inference is performed by a scalable Bayesian algorithm (EnKF/SM-EnKF or particle/Rao--Blackwellized filtering), and the output is summarized by posterior means $\mathbb{E}[u_k\mid y_{1:k}]$, uncertainty summaries (credible bands, spectra, probabilities of coherent structures), and predictive distributions for future states/observables. We evaluate performance using standard diagnostics and proper scoring rules, emphasizing calibration.

\subsection{Viability relative to deterministic solvers, EnKF, and 4DVar}

We emphasize two viability criteria for methodological Bayesian numerics. First, posterior uncertainty must be \emph{calibrated}: credible intervals should attain nominal coverage and predictive distributions should achieve good proper scores (e.g. log score or CRPS), rather than merely producing visually plausible bands. Second, the probabilistic layer must be \emph{computationally efficient} and preserve the scalability of modern PDE solvers; in practice, this means avoiding expensive adjoint/Hessian machinery solely for uncertainty quantification and avoiding full-state particle filtering in very high dimension.

We compare against three classical baselines: (i) a deterministic solver, i.e. time stepping of the discretized PDE producing a single trajectory with no uncertainty quantification; (ii) the ensemble Kalman filter (EnKF), which provides a scalable ensemble-based Gaussian approximation to the filtering distribution when observations are incorporated \citep{evensen2009, reich2015}; and (iii) variational 4DVar, which in observational settings typically produces a MAP estimator over a time window via iterative optimization with adjoints \citep{law2015}. While 4DVar is often highly effective for point estimation, uncertainty quantification is not automatic and usually requires additional approximations (ensembles of analyses or Hessian/low-rank constructions), increasing computational cost.

Our central claim is not that probabilistic methods outperform optimized point estimators on every metric, but that \emph{calibrated uncertainty at scale} can be achieved with a modest overhead over EnKF-like workflows. In particular, scale-mixture/NIG augmentations introduce low-dimensional latent variables that can be sampled (or approximated) cheaply, while retaining conditionally Gaussian updates and thus preserving the computational footprint of EnKF/Rao--Blackwellized filtering.

\begin{figure}[H]
\centering
\includegraphics[width=0.95\linewidth]{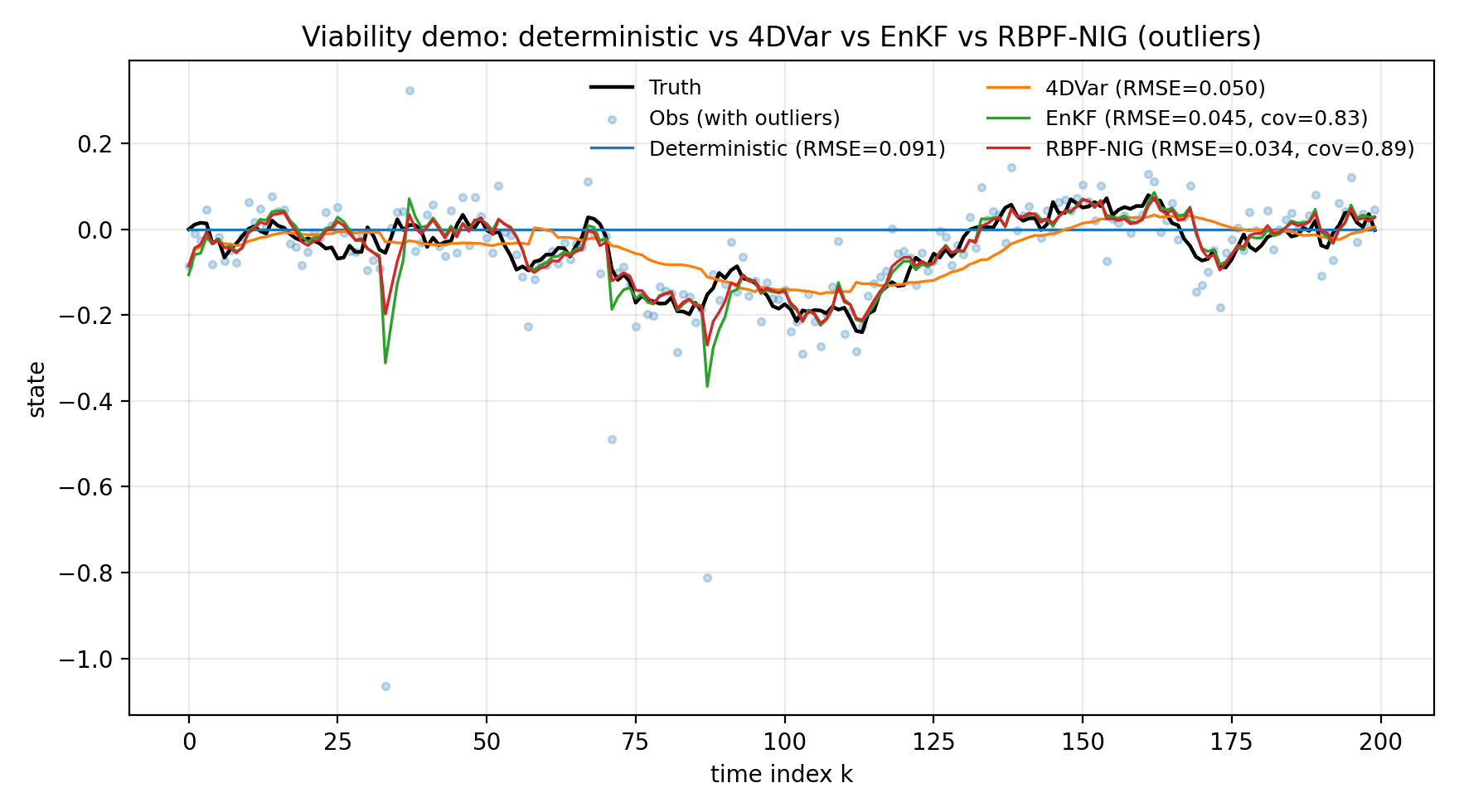}
\caption{Viability comparison on a 1D toy problem with outliers: deterministic forecast-only baseline, sliding-window 4DVar (MAP point estimate), EnKF (Gaussian uncertainty), and an NIG-motivated scale-augmented RBPF. The robust Bayesian filter adapts to outliers via latent scales and yields improved uncertainty calibration.}
\label{fig:viability-rbpf}
\end{figure}

Figure~\ref{fig:viability-rbpf} illustrates the key phenomenon on a simple linear state-space model with occasional observation outliers: Gaussian methods (EnKF/4DVar) provide reasonable point estimates but their uncertainty can be miscalibrated under outliers, while an NIG-motivated, scale-augmented Rao--Blackwellized particle filter improves uncertainty calibration (coverage) by adapting observation variance through latent scales. The script \texttt{code/demo\_viability\_baselines\_rbpf\_1d.py} reproduces the figure and summary metrics.

\subsection{Numerical Weather Prediction}

Operational centers solve primitive-equation models at extremely high dimension (often $>10^8$), and the incompressible Navier--Stokes viewpoint is most directly relevant in idealized regional/LES components and in the generic Bayesian methodology (state-space modeling, filtering, and uncertainty quantification).

Observations $y_k$ (radiances, winds, temperature, pressure, etc.) are related to the flow state through a (generally nonlinear) operator $h_k$:
\[
y_k = h_k(u_k) + \eta_k,
\]
where $\eta_k$ aggregates instrument noise and representativeness error. Linearizations lead to the form $y_k \approx H_k u_k + \eta_k$ used in ensemble filtering.

Operational weather centers employ ensemble Kalman filters with moderate ensemble sizes (often 10s--100s). Two numerically critical components are:

\textbf{Localization.} The sample covariance $\hat{P} = \frac{1}{N-1}\sum_i (u^i - \bar{u})(u^i - \bar{u})^T$ exhibits spurious correlations at large distances due to sampling noise. These are suppressed via Schur product with a compactly supported correlation function:
\begin{equation}
\hat{P}_{\text{loc}} = \rho \odot \hat{P},
\end{equation}
where $\rho_{ij}$ decays with distance $|x_i - x_j|$.

\textbf{Inflation.} Ensemble spread is multiplied by $\lambda > 1$ (typically 1.01--1.10) to compensate for underestimation of uncertainty.

Outliers and non-Gaussian innovations are common (e.g. cloudy radiances, intermittently biased sensors). The scale-mixture EnKF of Section~5 can be used as a robust alternative:
introduce latent scales $\tau_k$ (global, per-sensor, or per-observation-type) and run EnKF updates with inflated covariance $\tau_k R_k$. This preserves the EnKF computational footprint while yielding heavy-tailed robustness.

Beyond pointwise RMSE, a Bayesian implementation should report calibration diagnostics (innovation statistics, rank histograms for key observables, and coverage of credible intervals) to ensure that uncertainty quantification is meaningful.

\subsection{Turbulent Flow Control}

Closed-loop flow control requires real-time inference of an unobserved flow state from sparse sensors. A typical controlled, discretized model takes the form
\[
u_{k+1} = F_k(u_k,a_k) + \xi_k,\qquad y_k = h_k(u_k) + \eta_k,
\]
where $a_k$ is an actuator input (blowing/suction, plasma actuation, body forcing, etc.). The Bayesian filter produces the posterior $\pi_k(du)=\mathbb{P}(u_k\in du\mid y_{1:k},a_{1:k})$ used by the controller.

For moderate-dimensional reduced-order models, particle filters (possibly Rao--Blackwellized) can represent non-Gaussian posteriors induced by intermittency and actuator nonlinearities. For high-dimensional LES/DNS discretizations, EnKF variants are typically used, with localization tailored to the geometry (wall-bounded flows, jets, wakes).

Sensors in turbulent environments (e.g. wall shear stress, pressure taps) produce occasional bursts and outliers. NIG/scale-mixture observation models provide a principled way to avoid destabilizing the estimator: large residuals are absorbed by large inferred scales, preventing the update from over-correcting the state.

The posterior mean (or a low-dimensional posterior surrogate) can be fed into a receding-horizon controller (MPC) or linear-quadratic Gaussian (LQG) design. From a methodological standpoint, the key benefit is that the controller can account for uncertainty by penalizing risk or optimizing expected performance under the predictive distribution.

\section{Conclusion}\label{sec:conclusion}

We presented a methodological Bayesian viewpoint on numerical Navier--Stokes: discretized dynamics define a state-space model, and numerical solution becomes posterior computation. Uncertainty quantification follows immediately from the resulting posterior and predictive distributions, and when partial observations are available the same framework yields filtering as an additional capability. In 2D, stochastic representations provide intuition for Monte Carlo and particle approximations; in 3D, ensemble and particle-based methods scale Bayesian inference to large discretizations; and robust non-Gaussian error models can be incorporated via latent-variable augmentation without abandoning conditionally Gaussian updates.

Although the main goal of this paper is methodological rather than empirical, we included simple numerical experiments to illustrate two key computational points. First, we demonstrate that particle learning can perform online parameter learning without weight collapse by combining sufficient statistics (marginalization) with a resample--propagate (one-step smoothing) update (Figure~\ref{fig:pl-vs-sis}). Second, we run a small 2D Navier--Stokes pseudo-spectral solver in vorticity form and propagate a prior over uncertain viscosity and forcing amplitude through the time stepper using a particle/ensemble approximation. This produces a distribution over solution fields and quantities of interest (e.g. kinetic energy) without using any observations (Figure~\ref{fig:ns2d-solver-uq}). Finally, to illustrate robust observational updates when data are present, we consider a 1D linear system with occasional observation outliers: a Gaussian Kalman filter exhibits under-coverage, whereas a latent-scale (inverse-Gaussian) augmentation consistent with the NIG mixture yields a Rao--Blackwellized particle filter with improved uncertainty calibration and lower RMSE (Figure~\ref{fig:nig-rbpf}). A comparison against standard observational baselines (EnKF and windowed 4DVar) highlights a practical trade-off: robust and calibrated uncertainty under outliers can require an explicit non-Gaussian modeling layer (Figure~\ref{fig:viability-rbpf}). These experiments are fully reproducible via the accompanying scripts in \texttt{code/}.

The Bayesian framework is intended to be used selectively, matching algorithmic complexity to the problem regime.
For \textbf{pure forecasting} without observational updates or when uncertainty is not required, a deterministic solver remains appropriate.
When observations are incorporated and the regime is approximately Gaussian (moderate nonlinearity, well-characterized sensor errors), the \textbf{EnKF} offers a robust default that scales to very large state dimensions with localization and inflation.
When the priority is a \textbf{windowed MAP estimate} and adjoint infrastructure is available, \textbf{4DVar} is often competitive for point estimation; however, calibrated uncertainty typically requires additional approximations beyond the MAP.
Our proposed \textbf{scale-mixture / NIG-augmented filters} (SM-EnKF, RBPF over latent scales) are most valuable when observations exhibit \textbf{outliers, intermittency, or representativeness error} that violates Gaussian assumptions and when uncertainty calibration matters (e.g. probabilistic thresholds, risk-sensitive decisions, or downstream control).
Finally, full-state particle filtering is generally infeasible in very high dimension, but \textbf{Rao--Blackwellization and low-dimensional latent augmentation} can preserve tractability by restricting particles to the non-Gaussian degrees of freedom (scales and selected parameters).

Several directions would strengthen the bridge between theory, methodology, and practice. First, the most important next step is a \textbf{Navier--Stokes-scale empirical study} (e.g. 2D vorticity or 3D periodic box) comparing EnKF, 4DVar-style MAP, and scale-mixture variants under controlled non-Gaussian observation/forcing perturbations, reporting both accuracy and calibration (coverage, rank histograms, proper scores). Second, the NIG framework invites \textbf{online learning of non-Gaussian parameters} and adaptive modeling of representativeness error, including efficient samplers/approximations for the latent scales in very high dimension (global vs per-sensor/per-mode scales). Third, hybrid methods that combine \textbf{localization, low-rank structure, and particle updates} (particles over a small set of latent variables or reduced coordinates with EnKF-style Gaussian updates for the remainder) appear particularly promising for turbulence and flow-control settings. Finally, a systematic treatment of \textbf{computational budgets}---how to allocate ensemble size, localization radius, and augmentation complexity to meet calibration targets---would make the methodology more actionable for large-scale deployments.

\bibliography{ref}

\end{document}